# Iridium Enabled Field-free Spin-orbit Torque Switching of Perpendicular Magnetic Tunnel Junction Device

Yang Liu[1,3*], Bing Zhou[1,3], Zhengkun Dai[2,3], Enbo Zhang[2,3], Jian-Gang (Jimmy) Zhu[2,3]

[1] Department of Materials Science and Engineering, Carnegie Mellon University, 5000 Forbes Avenue, Pittsburgh, Pennsylvania, 15213, USA
[2] Department of Electrical and Computer Engineering, Carnegie Mellon University, 5000 Forbes Avenue, Pittsburgh, Pennsylvania, 15213, USA
[3] Data Storage Systems Center, Carnegie Mellon University, 5000 Forbes Avenue, Pittsburgh, Pennsylvania, 15213, USA

**Abstract**

Writing magnetic bits by spin-orbit torques (SOTs) arising from spin Hall effect creates new possibilities for ultrafast and low-power magnetoresistive random access memory (MRAM). For perpendicular MRAM, an extra in-plane field is required to break the symmetry for the deterministic SOT writing of the perpendicular storage layer. Although schemes have been demonstrated in external-field-free SOT switching of a perpendicular layer, practically integrating them with perpendicular MTJs still appears to be challenging. Here, we present experimental demonstration of spin-orbit torques (SOTs) switching a perpendicular magnetic tunnel junction (MTJ) device without applying an external magnetic field. An Ir layer is used to serve dual-purpose of both injecting the pure spin current via spin Hall effect and mediating an in-plane exchange field to the perpendicular free layer of the MTJ. Robust field-free SOT switching with pulsed write path current is demonstrated for various MTJ sizes ranging from 50 nm to 500 nm. The effect of MTJ size and pulse width on the critical switching current is studied. Combined micromagnetic simulations are carried out to provide in-depth analysis of the switching dynamics as well as the thermal effect on the switching.

**Introduction**

Over the last decades, magnetoresistive random access memory (MRAM) has come to the point where it's considered as a strongly competitive candidate for cache memory replacement[1–4]. In MRAM, storage units with perpendicular magnetic anisotropy (PMA) are preferred for that they render good thermal stability as well as high storage density[5–7]. Moreover, it has been shown that magnetization reversal via spin-orbit torques (SOT) arising from spin Hall effect can allow MRAM to be capable of ultra-fast and low-power writing operations[8–11]. To realize SOT induced magnetization switching in perpendicular MRAM, however, a static in-plane magnetic field is necessary for breaking the symmetry[9,12]. But applying a field externally is impractical for technological implementation.

Successful field-free perpendicular SOT switching has been realized with either utilizing antiferromagnetic materials[13–15], or incorporating an additional exchange coupling layer[16], or manipulating device geometry[17,18]. In those schemes, although field-free switching of a single Co/CoFe layer or Co/Ni multilayers is demonstrated, the integration with perpendicular MTJs doesn't seem to be so straightforward. The difficulties involve weak PMA and/or small exchange bias with FeCoB, inappropriate device design preventing MTJ integration and incapability to scale down.

In this work, we experimentally demonstrate the field-free SOT switching of perpendicular MTJs via utilizing Ir as the material of dual functions: generation of SOT along with mediating an in-plane field. The device adopts the usual three-terminal device design and doesn't require special engineering in device shape or inserting other functional layers. We show that the Ir-enabled field-free device exhibits reliable writing and reading operations, moving a step closer to the practical applications of SOT-related magnetoresistive devices.

In our approach, the key components for enabling zero-field switching is the [in-plane magnet/ Ir/ free perpendicular magnet] trilayer structure. The mechanism for the field-free switching is attributed to the interlayer exchange coupling and spin Hall effect of Ir[19]. Here, a preferable thickness of Ir is used so that the Ir layer is able to provide sufficient interlayer exchange coupling between adjacent ferromagnetic layers. As such, the free perpendicular layer ($m_p$ // z) above Ir experiences a local in-plane exchange coupling field ($H_x$ // x) from the lower in-plane magnetized layer ($m_i$ // x). During current-induced switching, charge current along x flows into the Ir layer, generating pure spin current with spin polarization $\sigma$ // y due to spin Hall effect. The spin current travelling along z is absorbed by the free layer $m_p$ and subsequently exert SOTs that can be decomposed into a Slonczewski-like torque $m_p \times (\sigma \times m_p)$ and field-like torque $m_p \times \sigma$. Facilitated by the local in-plane coupling field $H_x$, the SOTs are capable of driving the perpendicular magnetization reversal of $m_p$ in absence of an external magnetic field[19]. We also show that Ir fits well with FeCoB/MgO-based perpendicular MTJ and so the SOT driven magnetization reversal can be efficiently detected by the magnetoresistance change. In addition, the switching current threshold is characterized in terms of varying device size as well as pulse width, and in-depth study is carried out via micromagnetic simulations.

**Results**

We deposit the film at room temperature by magnetron sputtering with base pressure < 2 × $10^{-8}$ Torr. The film structure, as shown in Figure 1(a), is substrate/ IrMn(5)/ Co(2)/ Ru(0.85)/ Co(2)/ Ir(1.35)/ FeCoB(1.2)/ MgO(1.2)/ FeCoB(1.3)/ Ta(0.5)/ [Co(0.4)/ Pt(1)]$_{3.5}$/ Ru(0.85)/ [Co(0.4)/ Pt(1)]$_{6.5}$/ Pt(7) (in nm). The purpose of adopting an in-plane SAF instead of a single in-plane Co layer below Ir is to minimize the effect of the in-plane stray field on the perpendicular SOT switching. The thickness of the Ir layer corresponds to the second antiferromagnetic coupling peak in the Ruderman–Kittel–Kasuya–Yosida (RKKY) thickness dependence curve (see Supplementary Information). After deposition, the film is post-annealed at 300 $^0$C for 10 min with a 5000 Oe magnetic field applied along -x direction. Figure 1(a) also shows the hysteresis loops of the film, exhibiting the well-defined PMA of the MTJ.

The deposited film is processed into three-terminal magnetoresistive devices (Figure 1(b)) (more details in Method). The tunnel magnetoresistance ratio (TMR) is evaluated by measuring the resistance of MTJ with various perpendicular magnetic field. Figure 1(b) shows the TMR minor loop of 1.2-nm-thick free FeCoB layer of the device with a 150 nm-diameter MTJ. Bi-stable states of the low and high resistance of the MTJ is observed at zero field. The TMR is about 60%, with the abrupt magnetization reversal of the free layer.

In current-induced switching measurements, 200-ns current pulses are applied into the write path consisting of the Ir layer and the bottom pinned in-plane SAF structure. We connect a 10 MΩ resistor in series with the MTJ to reduce the bias voltage across the MgO layer. The resistance change of the MTJ is monitored by the lock-in technique. Figure 2(a) shows the well-defined current-induced switching of the MTJ at zero external field. Starting with a low-resistance state of MTJ and sweeping the current from positive to negative, a sharp increase in the MTJ resistance is observed when the current is below a negative threshold value, indicating the parallel (P) to antiparallel (AP) magnetization switching. Next, as the current is swept back to a sufficiently large positive value, the switching is reversed (AP to P switching), as shown by the abrupt drop in MTJ resistance. By applying a sequence of switching pulses (see Supplementary Information), the devices exhibit reliable zero-field switching capability, where the MTJ can be switched repeatedly between low and high resistance states with applied current of opposite polarities.

To further study the switching behaviors, an external in-plane field $H_x$ is applied during the current switching measurements. In Figure 2(b), when $H_x$ is applied along -x direction, we obtain almost identical switching behavior as the zero-field switching, i.e., the switching is clockwise. It's as expected since the exchange coupling field acting on the free FeCoB layer also lies in the -x direction due to the antiferromagnetic coupling characteristic of the 1.35 nm Ir. When the direction of $H_x$ is reversed, the switching loop becomes counterclockwise. It's due to that the anti-aligned $H_x$ now overcomes the local exchange coupling field (~100 Oe, see Supplementary Information) and thus the net field acting on the free FeCoB layer is along +x direction, leading to the opposite switching outcome.

Next, we compare the switching current for the devices with different sizes. In Figure 2(c), the results show that the switching current is linearly proportional to the width of the write path. The linear relationship can be interpreted by the fact that the SOT switching condition is determined by the current density instead of absolute current in the write path[8,12].

Furthermore, we study the effect of MTJ size on the switching current threshold. As shown in Figure 2(d), no significant change in the switching current is observed with MTJ sizes ranging from 500 nm to 150 nm while the switching current starts to increase as the MTJ size becomes smaller than 80 nm. To explain such trend in the switching current, we conduct micromagnetic simulations to compare the switching processes in devices with different sizes. Fig.3(a) illustrates the switching dynamics of the free FeCoB layer in a relatively large (240 nm x 240 nm) device. It's found that the switching starts with nucleation and formation of dense strips of reverse domains followed by rather complex expansion and merge of these reverse strip domains till the full reversal of the magnetic layer is reached. Note that such SOT resulted nucleation and expansion of reverse strip domains has been observed with utilization of Kerr microscopy in our previous experiments[19].

In Fig. 3(b), we show the initially formed reverse domains for five difference device sizes ranging from 300 x 300 nm$^2$ to 30 x 30 nm$^2$. Note the size of each reverse strip domain is essentially independent of device size. As the size of the free layer goes down, we observe the reducing number of domains. When the size is reduced below 60 nm, only a single reverse strip domain would nucleate under the SOT in the presence of the in-plane exchange field and thus near single domain switching is observed. Such transition from multi-domain switching in large devices to single domain switching in smaller devices causes a slight increase in the calculated switching current, as shown in Fig. 3(c). The simulated trend of the switching current as a function of MTJ size agrees well with our experimental data (Fig. 2(d)).

We then study the effect of pulse width on the switching current threshold. The duration of current pulse is varied from 20 ns to 5 $\mu$s and the effect on the critical switching current is plotted in Fig. 4. At current pulse width below 100 ns, the switching current value is essentially a linear function of inverse $\sqrt{\tau}$. When current pulse width is beyond 100 ns, the switching current levels off to almost a constant value.

Combined micromagnetic simulation using all measured properties of the fabricated MTJ stack implies that the SOT switching in our devices should occur within 1 ns. Furthermore, the simulation shows that when the pulse width becomes longer than 1 ns, the switching current threshold should become a constant value, no longer depending on the pulse width. However, the measured switching current dependence on pulse width ranging from 20 ns and 100 ns evidently disagree with the micromagnetic simulations. The most plausible reason should be that the perpendicular anisotropy of the free layer reduces due to the joule heating generated during the current pulse injection, further causing the change in the switching current threshold. Our simulation results shown in Fig. 5(a) indicate a linear relationship between the critical switching current and the perpendicular anisotropy of the free layer. Previous study has found that the perpendicular anisotropy of FeCoB/MgO decreases linearly with temperature, with a measured Blocking temperature at $T_B$ = 450K at which the anisotropy vanishes[20]. Assuming the same Blocking temperature for the free layer fabricated here, the temperature dependence of the interfacial perpendicular anisotropy of the free layer is:

$$K(T) = K_0 \left( 1 - \frac{T - T_{RT}}{T_B - T_{RT}} \right) \quad \text{Eq.(1)}$$

where $K_0$ is the perpendicular anisotropy constant at room temperature $T_{RT}$. With a current pulse of duration $\tau$, the temperature rise of the write wire underneath the free layer of the MTJ can be written as

$$T - T_{RT} = \frac{[I^2 R - k_{dissp}(T - T_{RT})] \cdot \tau}{C \cdot V} \quad \text{Eq(2)}$$

Where $k_{dissp}(T - T_{RT})$ is the power of heat dissipation, $CV_{effective}$ is the heat capacitance, $I$ is the current and $R$ is the resistance of the write path. Combining the two above equations with this linear dependence in Fig. 5(a), we have

$$I = \frac{I_o}{(K_0 - 2\pi M_S^2)} \left\{ K_0 \left( 1 - \frac{I^2 R}{\left(\frac{C \cdot V}{\tau} + k_{dissp}\right) \cdot (T_B - T_{RT})} \right) - 2\pi M_S^2 \right\} \quad \text{Eq.(3)}$$

where $I_0$ is the level-off switching current threshold at room temperature. The above equation should describe the switching current threshold as a function of pulse width $\tau$, with considering the anisotropy reduction of the free layer.

Fig. 5(b) combines micromagnetic simulation results (blue curve) and measurement data (red circles) as a function of inverse pulse width. The black curves are simulation results for the free layers with varying perpendicular anisotropy due to different joule heating, showing how the heat effect curve (blue curve) is generated. The results shown in Fig. 5(b) provides a good understanding on the current driven perpendicular SOT switching in our devices. If the current pulse width is shorter than 1 ns, the temperature of the write wire doesn't rise, and the free layer anisotropy remains unchanged. The switching current threshold follows the micromagnetic simulation result for the case with $K_0$. In this case, the switching occurs within a fraction of nanosecond and the threshold current will level off to a constant value for any pulse width longer than 1 ns. For the cases with current pulse width longer than 1 ns, the temperature elevates, and the perpendicular anisotropy of the free layer decreases accordingly. The measured data is actually the level-off current threshold at different anisotropy. A longer duration pulse (for $\tau < 100$ ns) leads to higher temperature, hence lower threshold current, until the heat generated becomes balanced by the heat dissipation. This lowest threshold current corresponding to the maximum temperature increase (from room temperature), which is 57 K in the case based on the calculation.

**Conclusion**

In summary, we have demonstrated reliable and repeatable field-free SOT switching of perpendicular MTJs, facilitated by a thin Ir layer which performs dual-function of injection of pure spin current via spin Hall effect and providing an in-plane field via exchange coupling. The switching dynamics has been investigated by both experiments and micromagnetic simulation. We found that the non-thermal impact SOT switching occurs over sub-nanosecond range since it is governed by magnetization precession. At nanosecond scale or longer, thermally induced anisotropy field degradation in the free layer of the MTJ becomes significant, resulting in lowering switching threshold current. Our approach for field-free switching is based on well-understood phenomena and doesn't require special works in order to integrate with magnetoresistive stacks.

The devices also exhibit potentials in ultra-fast switching speed and scaling down to dimensions of domain wall width. We believe such Ir-enabled manipulation of perpendicular MTJs can move forward the practical realizations of applications involving SOT memory, sensors and logic devices.

**Method**

All films are deposited by magnetron sputtering on Si/SiO$_2$ substrate at room temperature. We use DC sputtering to deposit metal layers and RF sputtering to deposit the MgO layer. The composition of IrMn is Ir$_{20}$Mn$_{80}$ (at%) and of FeCoB is Fe$_{17.5}$Co$_{52.5}$B$_{30}$ (at%). The films are post-annealed at 300$^0$C for 10 min with a 5000 Oe magnetic field applied along -x direction as shown in Figure 1.

To pattern three-terminal devices, we first deposit a hard mask of 60 nm C /5 nm SiN. Then we perform e-beam lithography with HSQ negative resist followed by reactive ion etching and ion beam etching to fabricate MTJ pillars. Next, we pattern the write path by photolithography with AZ4210 positive resist and etch away all layers using ion beam etching. After striping the resist, a layer of 140 nm SiN passivation layer is then deposited on the chip. We then do a low-angle ion beam etching to planarize the chip. Top and bottom lead channels are opened sequentially by photolithography with AZ4110 positive resist and reactive ion etching. At last, we fabricate the leads by photolithography, 5 nm Ti/ 80 nm Au deposition and lift-off process.

All electrical measurements are performed at room temperature. Before measurements, the devices are saturated with -10 kOe perpendicular field and relaxed back to zero field. This is to set the reference FeCoB layer as up magnetization state in all devices. The TMR loop is measured by detecting device resistance with sweeping perpendicular field. The current-induced switching loop is obtained by applying current pulses (with pulse width ranging from 5 $\mu$s to 20 ns) along the writing channel and measuring the device resistance after each pulse, with or without the presence of an in-plane field.

In the micromagnetic simulation, the Landau-Lifshitz-Glibert equation adding Slonczewski spin transfer torque

$$\frac{d\hat{m}_f}{dt} = -\gamma \hat{m}_f \times \vec{H} + \alpha\, \hat{m}_f \times \frac{d\hat{m}_f}{dt} + \frac{\hbar}{2e} \cdot \gamma \cdot \frac{J\theta_{SH}}{M_s \delta} \hat{m}_f \times \hat{p}_{SH} \times \hat{m}_f$$

is used to model the magnetization, unit vector $\hat{m}_f$, of the perpendicular magnetic free layer (for the MTJ) on top of the Ir layer. In the above equation, $\hat{p}_{SH}$ is the unit vector of injected spin polarization, $M_s\delta$ is the magnetic moment density of the free layer, $J$ is the current density in the Ir layer and $\theta_{SH}$ is the spin Hall angle. $\vec{H}$ is the effective magnetic field defined as

$$\vec{H} = -\frac{1}{M_S} \frac{\partial E}{\partial \hat{m}}$$

where the energy here includes the perpendicular anisotropy energy, the exchange energy arising from spatial inhomogeneity of the magnetization, and the coupling energy between the free layer and the in-plane magnetic layer in the SAF underneath the Ir layer:

$$E = K\left[1 - (\hat{k} \cdot \hat{m})^2\right] + A\left[(\nabla m_x)^2 + (\nabla m_y)^2 + (\nabla m_z)^2\right] + \sigma\, \hat{m}_f \cdot \hat{m}_1$$

where $A = 1.0 \times 10^{-6}$ erg/cm is the exchange stiffness constant used and 0.014 erg/cm² is used for the interlayer exchange coupling between the magnetic layers on either side of the Ir layer.

The two in-plane magnetic layers in the synthetic antiferromagnet, unit magnetization vectors $\hat{m}_1$ and $\hat{m}_2$, are also modeled simultaneously with the following LLG equations:

$$\frac{d\hat{m}_{1,2}}{dt} = -\gamma \hat{m}_{1,2} \times \vec{H}_{1,2} + \alpha\, \hat{m}_{1,2} \times \frac{d\hat{m}_{1,2}}{dt}$$

Each of the three magnetic layers is meshed into a 2D array of mesh elements of 2 nm x 2 nm thickness in size. The magnetization vectors of the mesh cells for all the elements are solved simultaneously with numerical integrations of the above coupled dynamic equations.

**Reference**

1. Wolf, S. A. and Awschalom, D. D. and Buhrman, R. A. and Daughton, J. M. and von Molnr, S. and Roukes, M. L. and Chtchelkanova, A. Y. and Treger, D. M. Spintronics: a spin-based electronics vision for the future. *Science* **294**, 1488–95 (2001).

2. Huai, Y. Spin-transfer torque MRAM (STT-MRAM): Challenges and prospects. *AAPPS Bull.* **18**, 33–40 (2008).

3. Chen, E. *et al.* Advances and future prospects of spin-transfer torque random access memory. *IEEE Trans. Magn.* **46**, 1873–1878 (2010).

4. Kent, A. D. & Worledge, D. C. A new spin on magnetic memories. *Nat. Nanotechnol.* **10**, 187–191 (2015).

5. Ikeda, S. *et al.* A perpendicular-anisotropy CoFeB-MgO magnetic tunnel junction. *Nat. Mater.* **9**, 721–724 (2010).

6. Zhu, J.-G. G. J. & Park, C. D. Magnetic tunnel junctions. *Mater. Today* **9**, 36 (2006).

7. Sato, H. *et al.* Perpendicular-anisotropy CoFeB-MgO magnetic tunnel junctions with a MgO/CoFeB/Ta/CoFeB/MgO recording structure. *Appl. Phys. Lett.* **101**, (2012).

8. Liu, L. Spin-torque switching with the giant spin Hall effect of tantalum. *Science (80-. ).* **336**, 555–558 (2012).

9. Liu, L., Lee, O. J., Gudmundsen, T. J., Ralph, D. C. & Buhrman, R. A. Current-induced switching of perpendicularly magnetized magnetic layers using spin torque from the spin Hall effect. *Phys. Rev. Lett.* **109**, 96602 (2012).

10. Pai, C. F. *et al.* Spin transfer torque devices utilizing the giant spin Hall effect of tungsten. *Appl. Phys. Lett.* **101**, 1–5 (2012).

11. Cubukcu, M. *et al.* Spin-orbit torque magnetization switching of a three-terminal

perpendicular magnetic tunnel junction. *Appl. Phys. Lett.* **104**, (2014).

12. Fukami, S., Anekawa, T., Zhang, C. & Ohno, H. A spin–orbit torque switching scheme with collinear magnetic easy axis and current configuration. *Nat. Nanotechnol.* 1–27 (2016). doi:10.1038/nnano.2016.29

13. Fukami, S., Zhang, C., DuttaGupta, S., Kurenkov, A. & Ohno, H. Magnetization switching by spin–orbit torque in an antiferromagnet–ferromagnet bilayer system. *Nat. Mater.* **15**, 535 (2016).

14. van den Brink, A. *et al.* Field-free magnetization reversal by spin-Hall effect and exchange bias. *Nat. Commun.* **7**, 10854 (2016).

15. Oh, Y.-W. *et al.* Field-free switching of perpendicular magnetization through spin–orbit torque in antiferromagnet/ferromagnet/oxide structures. *Nat. Nanotechnol.* **11**, 878 (2016).

16. Lau, Y.-C., Betto, D., Rode, K., Coey, J. M. D. & Stamenov, P. Spin–orbit torque switching without an external field using interlayer exchange coupling. *Nat. Nanotechnol.* **11**, nnano-2016 (2016).

17. Yu, G. *et al.* spin – orbit torques in the absence of external magnetic fields. **9**, (2014).

18. You, L. *et al.* Switching of perpendicularly polarized nanomagnets with spin orbit torque without an external magnetic field by engineering a tilted anisotropy. *Proc. Natl. Acad. Sci.* **112**, 10310–10315 (2015).

19. Liu, Y., Zhou, B. & Zhu, J. J. Field-free Magnetization Switching by Utilizing the Spin Hall Effect and Interlayer Exchange Coupling of Iridium. *Sci. Rep.* 1–7 (2019). doi:10.1038/s41598-018-37586-4

20. Lee, K.-M., Choi, J. W., Sok, J. & Min, B.-C. Temperature dependence of the interfacial magnetic anisotropy in W/CoFeB/MgO. *AIP Adv.* **7**, 65107 (2017).

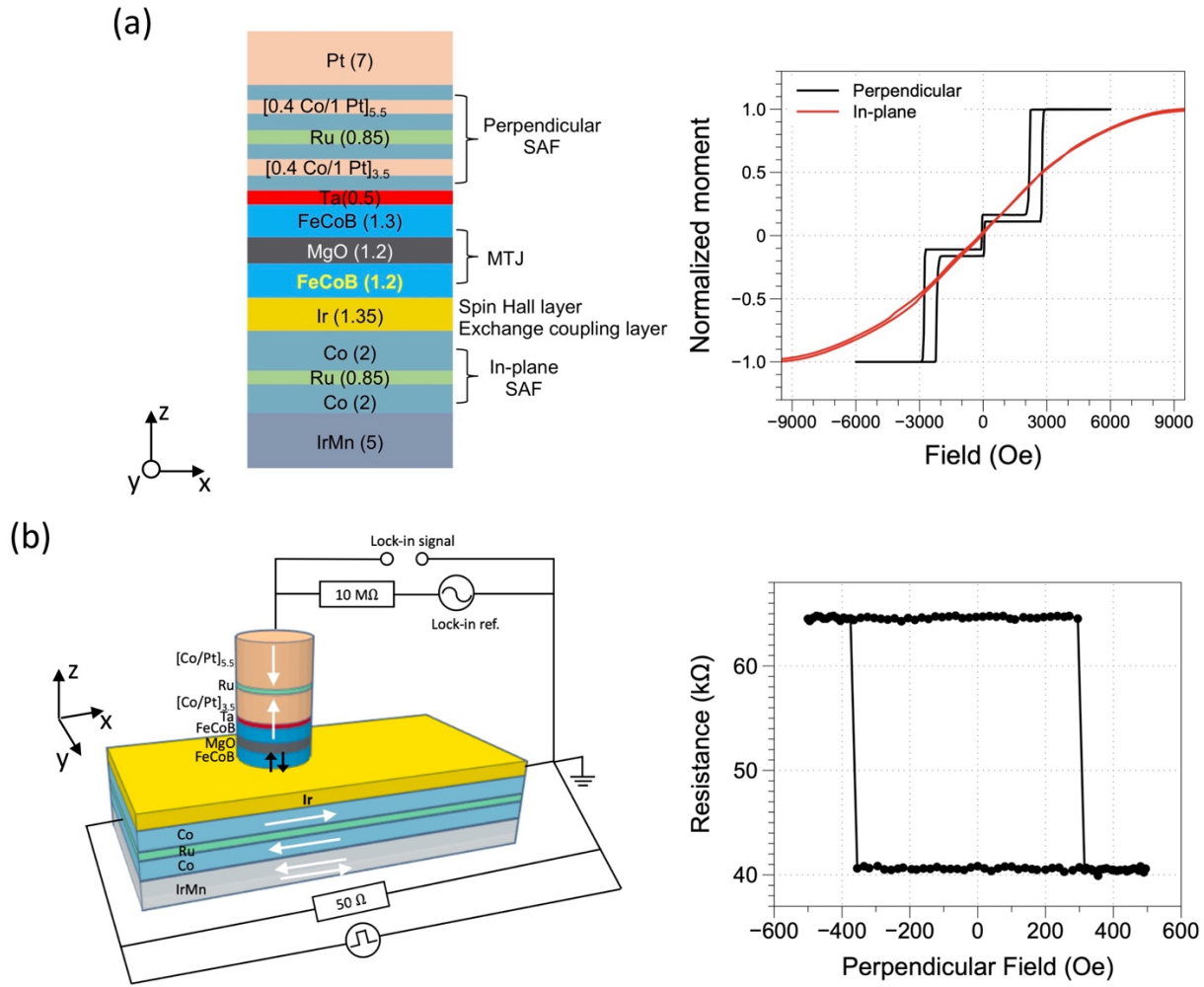

Figure 1. (a) Left: film stack, unit in nanometer. Right: M-H loops of the film. (b) Left: schematic of the device and testing set-up. Right: Perpendicular-field-driven TMR minor loop of a device with a 150-nm-diameter MTJ.

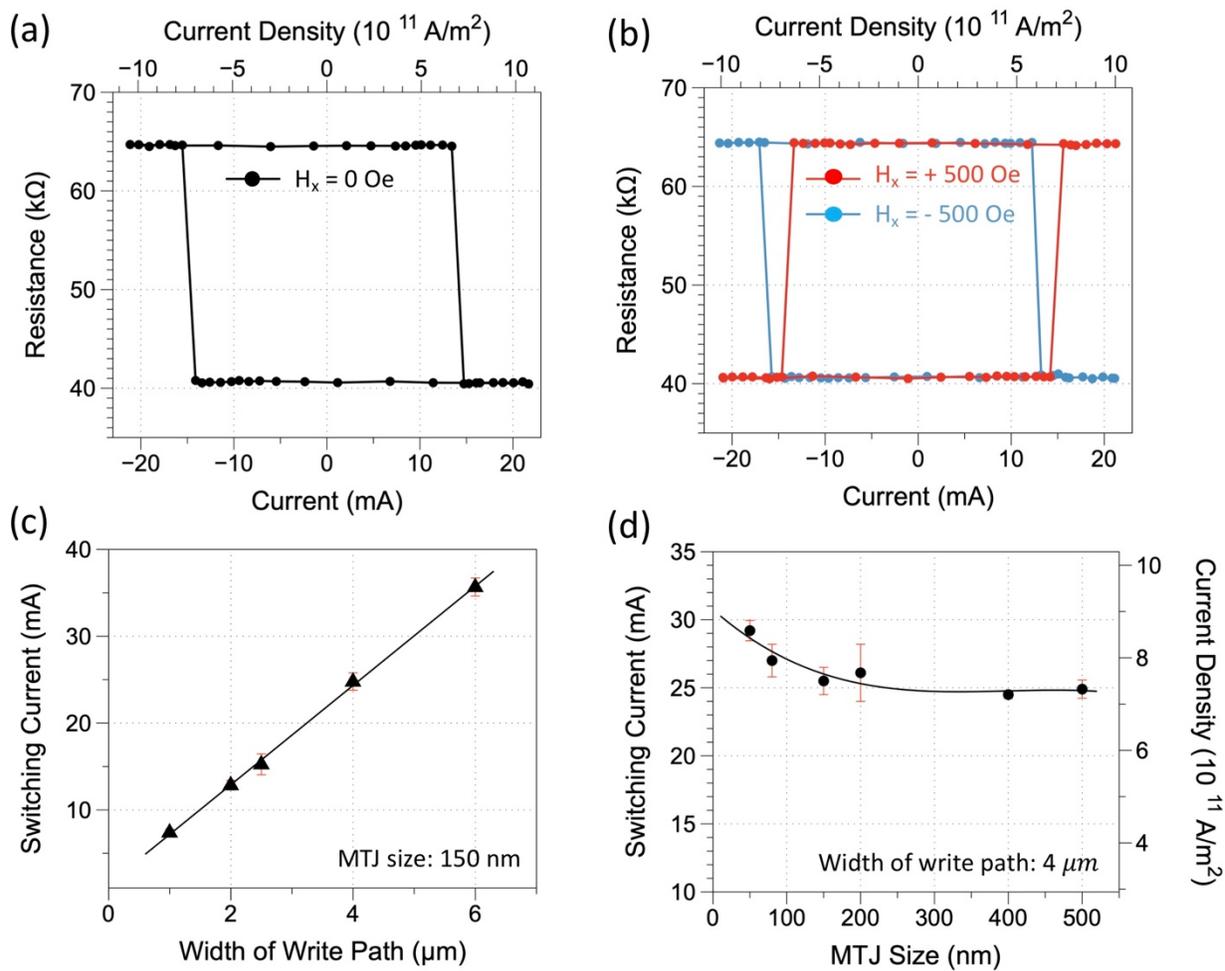

Figure 2. Upper: Current-induced switching results (a) at zero external field and (b) with an externally applied field along current direction. The device has a 2.5 μm-width write path and a 150-nm-diameter MTJ. Lower: Switching current as a function of (c) the width of the write path and (d) the MTJ size.

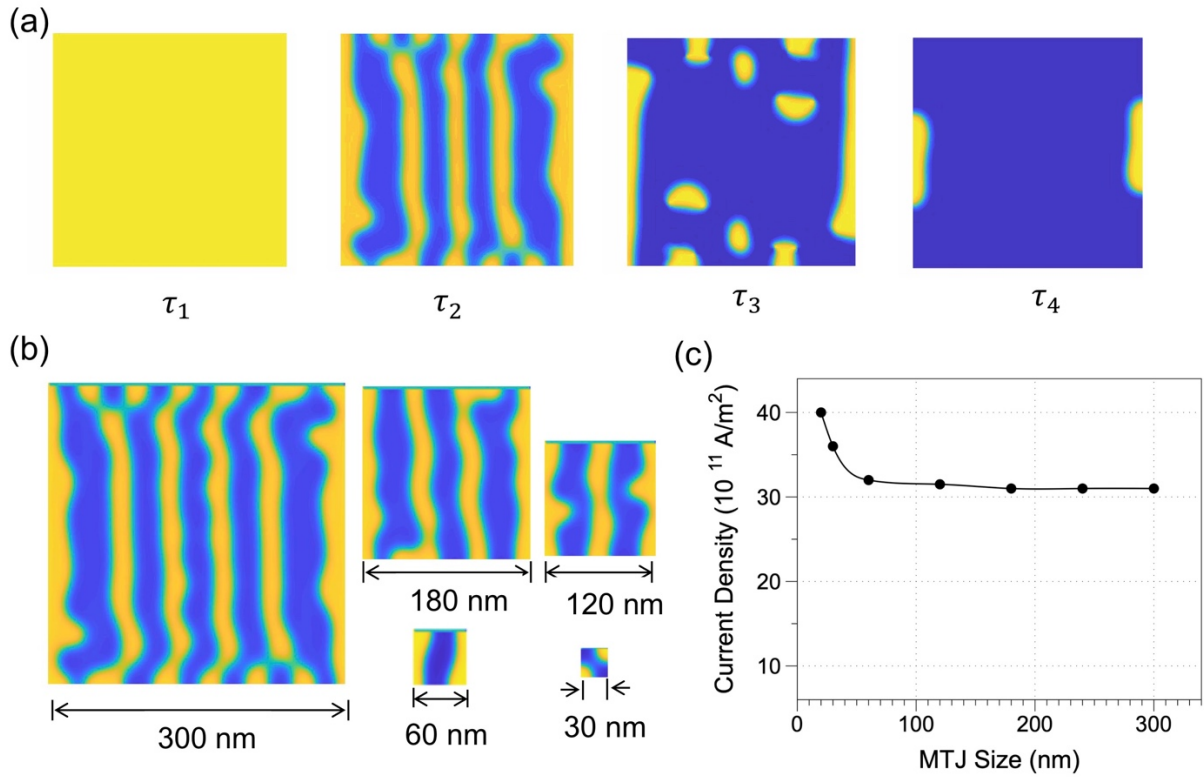

Figure 3. Micromagnetic simulation of the zero-field switching dynamics. (a) Switching dynamics of the free FeCoB layer in a 240 nm x 240 nm device. Time: $\tau_1 < \tau_2 < \tau_3 < \tau_4$. (b) Comparison of the switching in devices with various sizes. (c) Simulated trend of the switching current as a function of MTJ size.

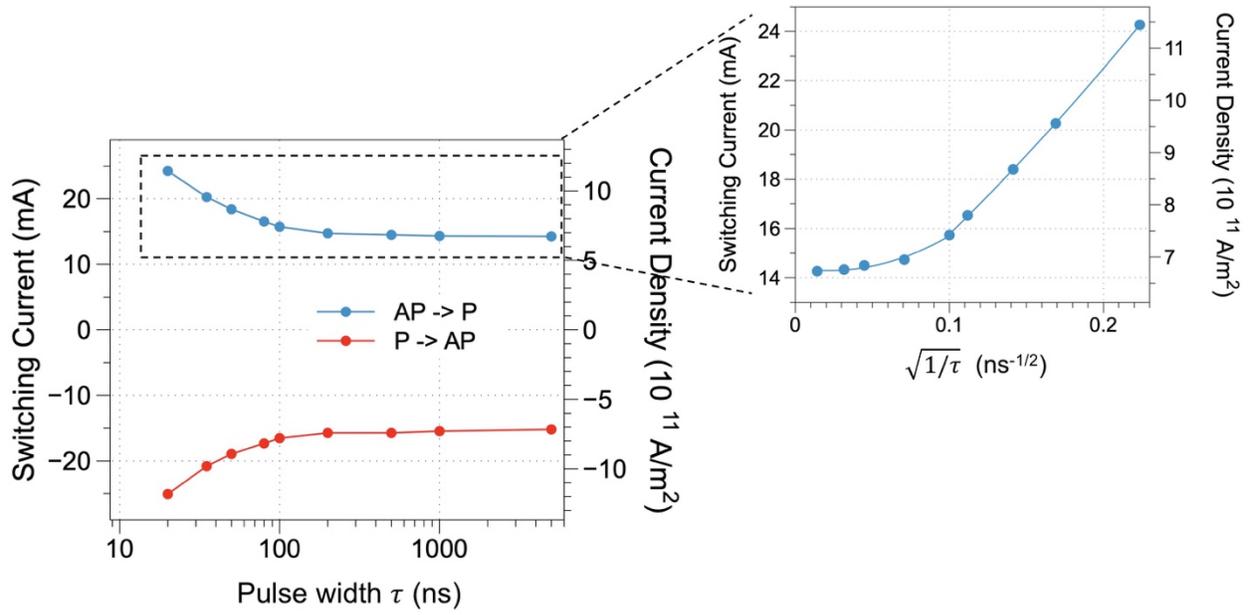

Figure 4. Switching current with different pulse width $\tau$. Upper right: Switching current is plotted as a function of inverse $\sqrt{\tau}$.

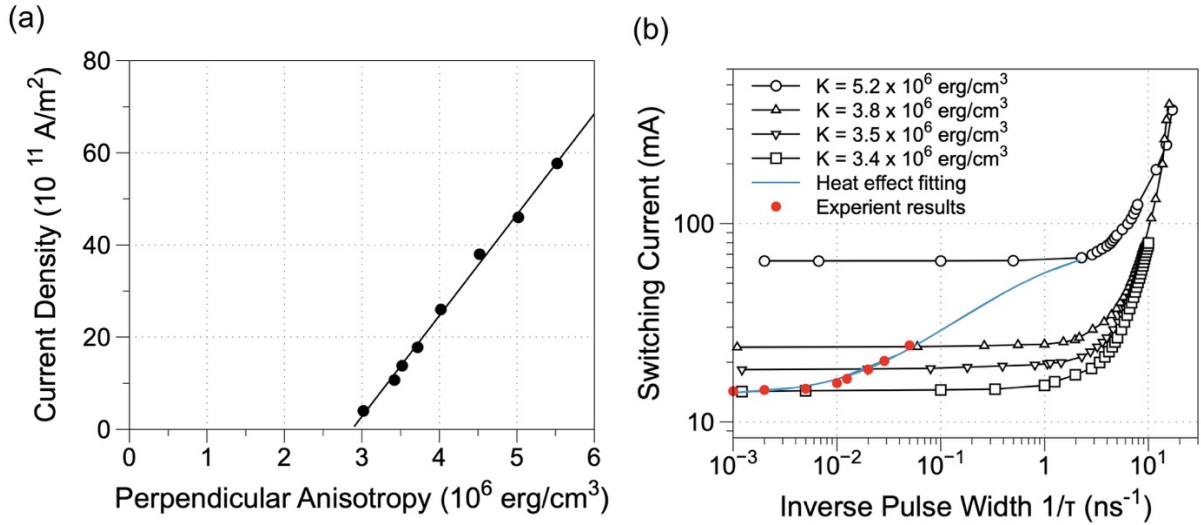

Figure 5. (a) Simulated critical current density as a function of the perpendicular anisotropy of the free layer. (b) Simulated switching current as function of the pulse width with different perpendicular anisotropy of the free layer (black curves). The blue curve fits the level-off current of each black curve and shows how the switching current changes when considering the heat effect. The red dots correspond to the experimental data.

# Supplementary Information

## 1. Interlayer exchange coupling of Iridium (Ir)

Film stacks for characterize the interlayer exchange coupling of Ir are substrate/ Co (2 nm)/ Ir ($t_{Ir}$)/ Co (2 nm)/ Ta (2 nm) where $t_{Ir}$ ranges from 0.6 nm to 1.65 nm. In-plane hysteresis loops are measured by alternating gradient field magnetometer (AGFM) and used to determine the strength of interlayer exchange coupling. Figure S1 shows the exchange coupling field as a function of Ir layer thickness. Note that only antiferromagnetic coupling is shown in Figure S1. As can be seen, the first antiferromagnetic coupling peak locates at 0.6 nm and the second at 1.35 nm. In the field-free switching experiments, we choose Ir layer thickess to be 1.35 nm because the exchange coupling via 0.6 nm Ir is so strong that the perpendicular layer can be pulled into in-plane direction.

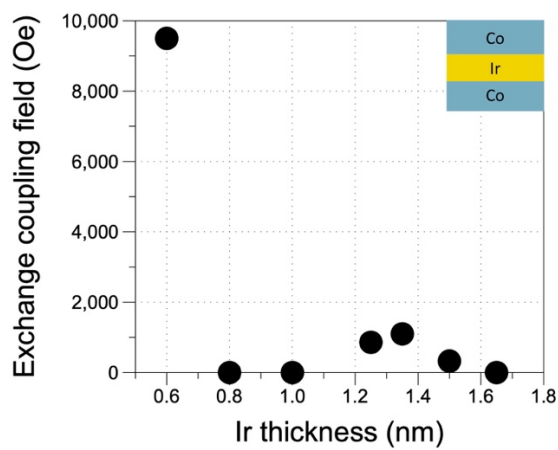

Figure S1. Antiferromagnetic coupling strength via an Ir spacer with various Ir layer thickness

## 2. Field-free SOT switching of a perpendicular FeCoB layer

Here we show the field-free SOT switching of a perpendicular FeCoB layer. The film stack is substrate/ $Ir_{20}Mn_{80}$(5 nm)/ Co(2 nm)/ Ru(0.85 nm)/ Co(2 nm)/ Ir (1.35 nm)/ FeCoB (1.2 nm)/ MgO (2 nm)/ Ta capping. The film is post-annealed at 300°C for 10 min with a 5000 Oe in-plane field. The perpendicular M-H loop in Figure S2(b) exhibits the strong perpendicular magnetic anisotropy of the FeCoB layer.

Then we pattern the film into cross-shape Hall-bar devices with 1 $\mu m$-wide current-channel and 1 $\mu m$-wide voltage-channel. During the switching measurements, an external in-plane field is applied along the current direction with the magnitude ranging from 500 Oe to -500 Oe. In Figure S2(c), we observe well-defined and complete SOT switching of the perpendicular FeCoB layer at zero field. Such field-free switching, as discussed in the main manuscript, is attributed to the interlayer exchange coupling as well as the spin Hall effect of Ir. It's also observed that the switching loop collapses at a 100 Oe external in-plane field. No switching happens because the anti-aligned external field cancels with the local exchange coupling field, indicating the coupling field acting on the FeCoB layer has the magnitude of about 100 Oe.

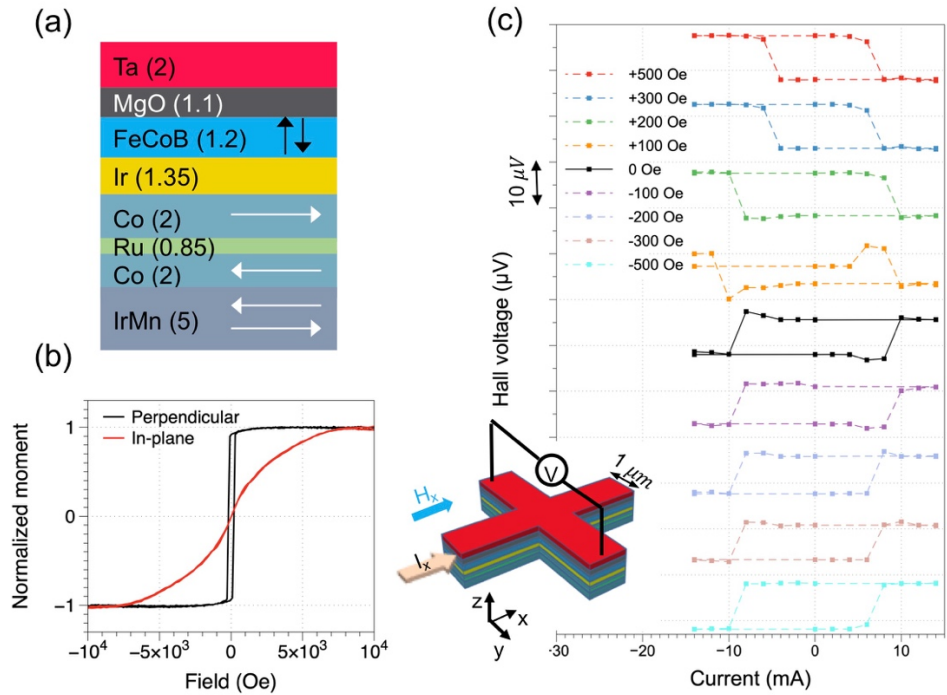

Figure S2. (a) Film stack (unit in nm). (b) M-H loops of the film. (c) Current-induced magnetization switching w/o an external field.

**3. Zero-field switching of perpendicular MTJs with switching pulse cycles**

Figure S3 shows the detected signals from MTJ resistance change with a sequence of positive and negative switching-current pulses. As can be seen, the MTJ can be switched back and forth between low and high resistance states by applying current of opposite polarities. This is to demonstrate that the Ir-enabled field-free switching devices exhibit reliable writing and reading performance.

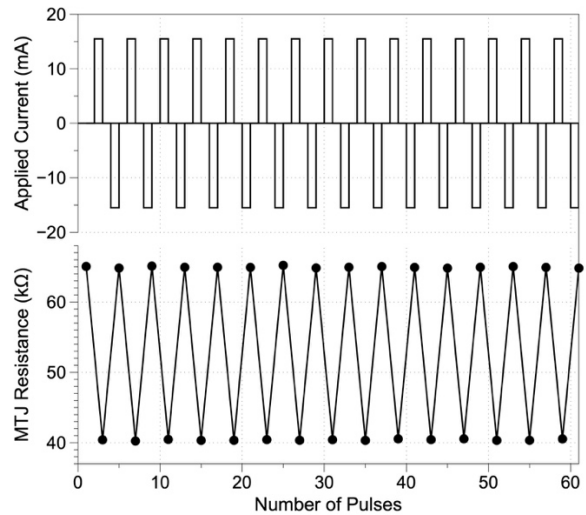

Figure S3. Field-free switching of perpendicular MTJs with a sequence of positive and negative switching pulses.